\documentclass{PoS}

\usepackage{cite}

\newcommand{\be}{\begin{equation}}
\newcommand{\ee}{\end{equation}}
\newcommand{\bra}{\langle}
\newcommand{\ket}{\rangle}
\newcommand{\Tr}{\mbox{Tr}}

\title{Recent developments at finite density on the lattice}

\ShortTitle{Recent developments at finite density on the lattice}

\author{\speaker{Gert Aarts} \\
        Department of Physics, College of Science, Swansea University,
        Swansea, United Kingdom\\
        E-mail: \email{g.aarts@swan.ac.uk}}

\abstract{
Some recent developments to handle the numerical sign problem in QCD and related theories at nonzero density are reviewed. In this contribution I focus on changing the integration order to soften the severity of the sign problem, the density of states, and the extension into the complex plane (complex Langevin dynamics and Lefshetz thimbles).
}

\FullConference{9th International Workshop on Critical Point and Onset of
Deconfinement\\
                 17-21 November, 2014\\
                 ZiF (Center of Interdisciplinary Research), University of Bielefeld, Germany}

\begin{document}

\section{Introduction}

The QCD partition function on the lattice is usually written as
\be
 Z = \int DU D\bar\psi D\psi\, e^{-S} = \int DU\, e^{-S_{\rm YM}} \det M,
 \ee
  where $U$ denote the gauge links, $S_{\rm YM}$ is the Yang-Mills action, and in the final expression the quark fields have been integrated out to yield the fermion determinant. 
At nonzero quark chemical potential $\mu$, this determinant is complex,
\be
 [\det M(\mu)]^* = \det M(-\mu^*),
 \ee
 unless the chemical potential is taken as purely imaginary. With a complex Boltzmann weight  (for real $\mu$), numerical lattice QCD simulations relying on importance sampling are not  straightforward and hence the QCD phase diagram has not yet been determined from first principles. This problem is usually referred to as the sign problem and may refer more generally to any problem in which a complex Boltzmann weight is encountered or in which important minus signs appear due to the Grassmann nature of the fermionic fields.
 
 As is well-known \cite{deForcrand:2010ys}, the problem of a complex  weight is hard and cannot be ignored. For instance, simply taking the absolute value of the determinant and incorporating the phase factor later (reweighting) will typically destroy the correct physics, especially in the thermodynamic limit (overlap problem).
 In this contribution I will present a selection of recent approaches which aim to tackle the sign problem in QCD and related theories. Some complementary reviews touching on related developments can be found in Refs.\ \cite{Aarts:2013bla,Aarts:2013uxa,Chandrasekharan:2013rpa,Ejiri:2013lia,Aarts:2013naa,Gattringer:2014cpa,Sexty:2014dxa} and I will attempt to avoid too much repetition. In the next section, we start with the notion that  the overlap problem might be milder if  the order of integration is changed. Section \ref{sec:dos} is devoted to the density of states, in which the integration is carried out in two steps. Finally, in section \ref{sec:complex} the complex nature of the weight is taken seriously and the configuration space is extended into the complex plane, using either complex Langevin dynamics or integration along Lefschetz thimbles.

\section{Change integration order}
\label{sec:order}

In the standard formulation of lattice QCD, the sign problem arises due to the complex fermion determinant at nonzero chemical potential, obtained by performing the Grassmann integrals over the quark fields. Hence it makes sense to not integrate out the fermions immediately but instead perform the integral over the gauge links first. This is of course easier said than done, since the gauge sector is an interacting theory by itself. The first step is to consider the strong-coupling limit, with the gauge coupling $\beta=2N_c/g^2=0$. In this case the link integrals factorise to one-link integrals, which can be done analytically. The resulting remaining partition function is known as the Monomer-Dimer-Polymer (MDP) system \cite{Karsch:1988zx} and describes worldlines of mesons and baryons. In recent years this system has been studied in more detail \cite{deForcrand:2009dh,Unger:2011it}, which has been possible mostly due to the use of new worm-type algorithms.

The obvious question is how to go beyond the strong-coupling limit. Here recent progress has been made \cite{deForcrand:2014tha} by including the first ${\cal O}(\beta)$ corrections, namely by expanding the Yang-Mills weight as
\be
e^{-S_{\rm YM}} = 1+\beta [\mbox{plaquettes}] + \ldots
\ee
The  ${\cal O}(\beta)$ corrections can be included either by reweighting from the $\beta=0$ ensemble or by again using worm-type algorithms. Details at this meeting were presented by Unger.

\begin{figure}[t]
\begin{center}
 \includegraphics[height=8cm]{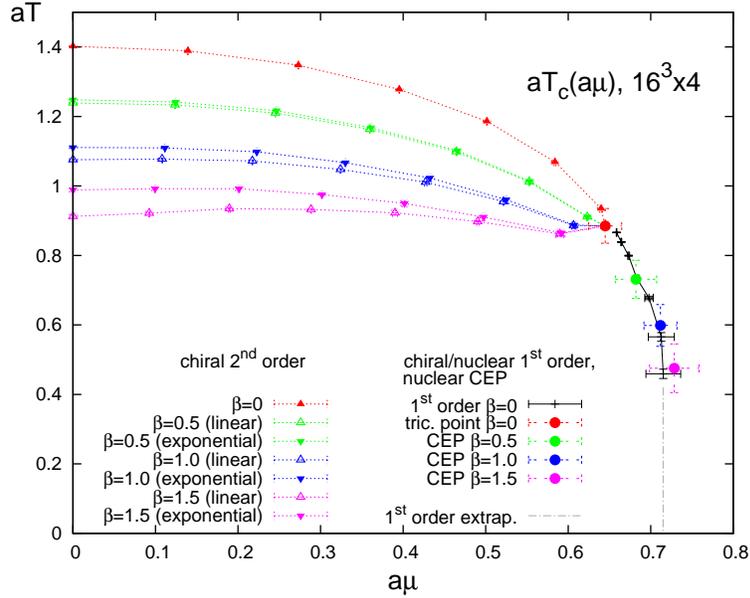} 
\caption{Phase diagram in the $aT-a\mu$ plane in the strong-coupling limit ($\beta=0$) and by including  ${\cal O}(\beta)$ corrections, in the chiral limit. From Ref.\ \cite{deForcrand:2014tha}. }
\label{fig:sc}
\end{center}
\end{figure}
\begin{figure}[t]
\begin{center}
  \includegraphics[height=4cm]{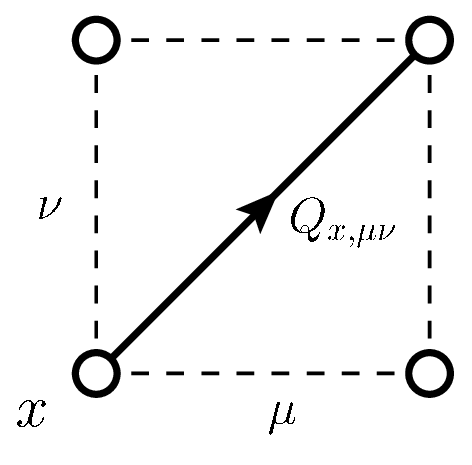}  \hspace*{1cm}
  \includegraphics[height=4cm]{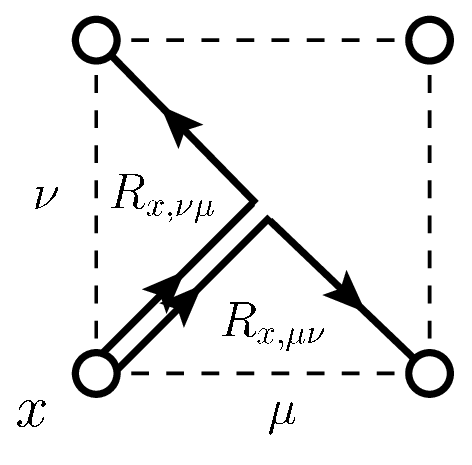}
\caption{Illustration of how auxiliary fields $Q$ and $R$ reduce a four-link plaquette to decoupled single links in the presence of auxiliary fields. From Ref.\ \cite{Vairinhos:2014uxa}.  }
\label{fig:hs}
\end{center}
\end{figure}

An example of the phase diagram is shown in Fig.\ \ref{fig:sc}, in the $T-\mu$ plane (in lattice units), for massless quarks. For small chemical potential, the transition is second order, while for larger chemical potential it is first order. The second and first-order line meet at a tricritical point. It can be seen that the critical temperature $aT_c$ decreases with increasing $\beta$, whereas the tricritical point only depends weakly on $\beta$. On the other hand, at nonzero $\beta$, the critical endpoint of the nuclear transition, studied via the baryon density, shifts to smaller temperature.
Clearly the next step is to include ${\cal O}(\beta^2)$ corrections, in order to see whether the expansion converges in any practical sense.

Another approach to go beyond the strong-coupling limit is to include the plaquettes by integrating them in steps. Each plaquette consists of four links and the idea here is to decouple those by introducing sets of auxiliary fields, as in a Hubbard-Stratonovich transformation.
This is illustrated in Fig.\ \ref{fig:hs}: in the figure on the left the auxiliary $Q$ fields breaks down the plaquette into two two-link components, while on the right the fields $R$ reduce those to decoupled single links. 
This setup was originally discussed 30 years ago \cite{Fabricius:1984wp}
 and has been recently taken up by Vairinhos \cite{Vairinhos:2010ha} and de Forcrand and Vairinhos \cite{Vairinhos:2014uxa}. 
While the link integrals can now be done, the integrals over the auxiliary fields of course remain. It is clear that this approach leads to a completely alternative representation of the QCD partition function and hence there is a hope that the sign problem may appear in a milder form. For an alternative formulation, see Ref.\ \cite{Brandt:2014rca}.
Recent developments were presented at this meeting by Vairinhos.

\section{Density of states}
\label{sec:dos}

The basic idea in the approach known as density of states is to do the path integral,
\be
Z = \int DU\, w(U),
\ee
in two steps, using contrained simulations. For instance, if we consider the density of states for an operator $x$, defined as
\be
\rho(x) = \int DU\, w(U)\delta[x-x(U)],
\ee
observables depending on $x$ can be reconstructed via
\be
\bra O(x) \ket = \frac{\int dx\, \rho(x) O(x)}{\int dx\, \rho(x)}.
\ee
Variations on this idea have been around for a long time, and are known as the  histogram method, factorisation, Wang-Landau, etc 
\cite{Gocksch:1988iz,wl,Ambjorn:2002pz,Fodor:2007vv,Ejiri:2013lia}.

\begin{figure}[t]
\begin{center}
  \includegraphics[height=7cm]{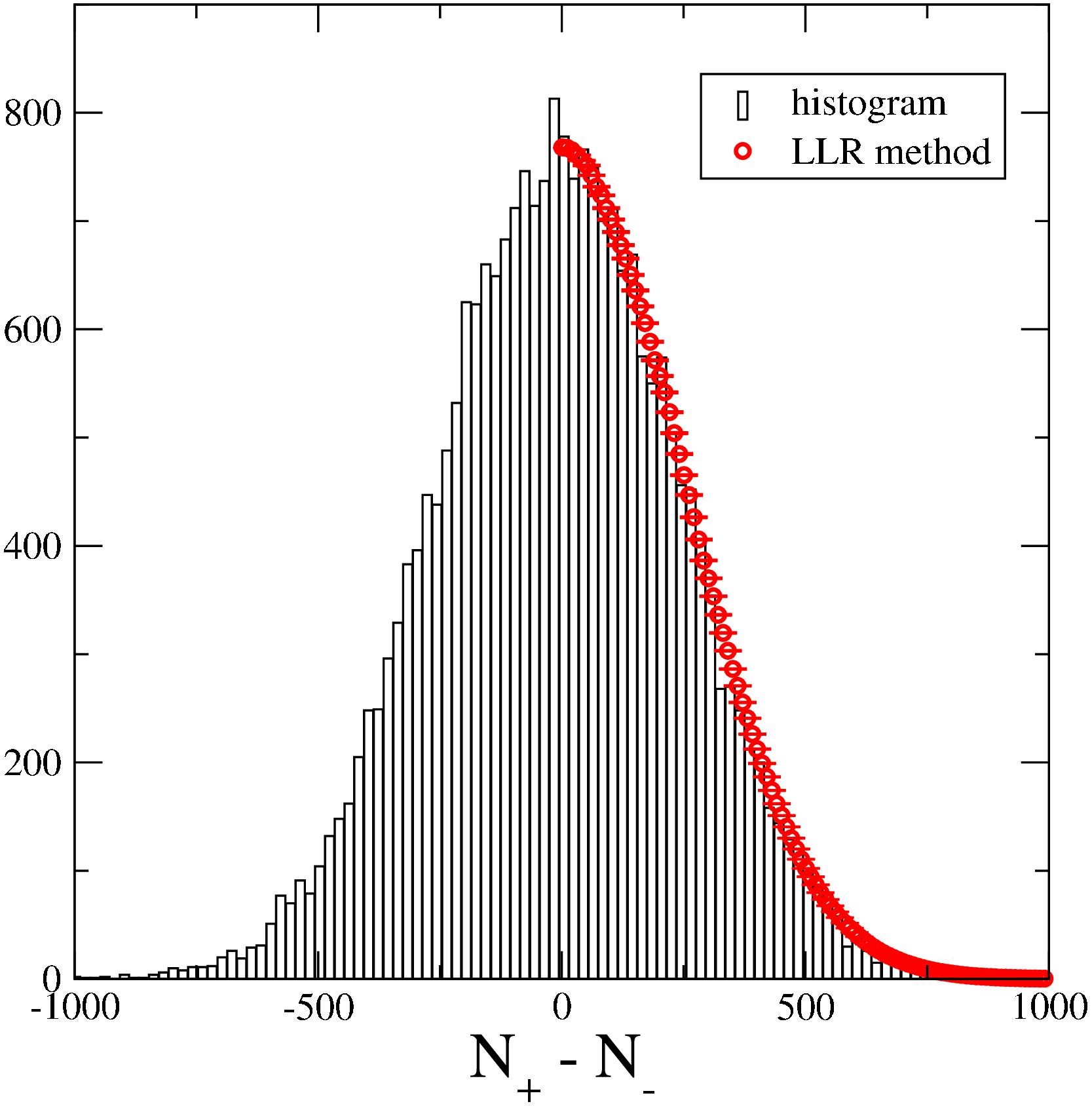}
  \includegraphics[height=7cm]{deltaN_histo_log2.eps}
\caption{Density of states in the Z(3) spin model at nonzero chemical potential, as  a function of $x=N_+-N_-$, the net density, on a normal scale (left) and on a logarithmic scale (right). Also shown is the histogram obtained with ordinary sampling. Infrequent events, say with $|x|\gtrsim 1000$, are not covered by the histogram but can be sampled with the density of states. From Ref.\ \cite{Langfeld:2014nta}.
  }
\label{fig:dos1}
\end{center}
\end{figure}

The main issues in this approach are quite obvious from the expressions above: 
\begin{enumerate}
\item the contrained integral should have a positive weight, so that it can be determined unambiguously;
\item  the weight $\rho(x)$ should be computable to very high {\em relative} precision.
\end{enumerate}
  Let us now consider theories with a complex weight, written as
\be
w(U) = |w(U)| e^{i\theta}.
\ee
As observable we consider the density $n$ and for simplicity we assume that the phase factor depends  only on the density, i.e.\ $\theta=\theta(n)$. 
The positive density of  states is then
\be
\rho(x) = \int DU\, |w(U)|\delta[x-n(U)],
\ee
and observables are given by
\be
\bra O(x) \ket = \frac{1}{Z} \int dx\, \rho(x) e^{i\theta(x)} O(x),
\quad\quad\quad\quad
Z = \int dx\, \rho(x) e^{i\theta(x)}.
\ee
A recent promising reincarnation of the density of states has been introduced by Langfeld, Lucini and Rago, and dubbed {\em local linear relaxation} or LLR \cite{Langfeld:2012ah}.
There are various improvements compared to previous work:
\begin{enumerate}
\item precise sampling of $\rho(x)$ over many orders of magnitude;
\item forced sampling in bins of width $\Delta x$ around $x$ with relative fluctuations of $\lesssim 1\%$ and no loss of efficiency;
\item precise integration over the oscillating function $\rho(x) e^{i\theta(x)}$.
\end{enumerate}
An application to finite density was recently given in the Z(3) spin model with a nonzero chemical potential in Refs.\ \cite{Langfeld:2014nta,Mercado:2014dva,Lucini:2014wga}.
The result for the density of states for $x=N_+-N_-$, the net particle density, is shown in Fig.\ \ref{fig:dos1}, on  a normal scale (left) and on a logarithmic scale (right). Also shown is the histogram, constructed by simply recording values of the density obtained during a standard simulation. Here it is clear that {\em rare} events, say with $|x|\gtrsim 1000$, occur so infrequently that they are not covered by the standard histogram. On the other hand, the density of states extends over more than 60 orders of magnitude, with $|x|$ up to 5500, since sampling in bins around a given value of $x$ is enforced. Such precision and extent is required to perform the integral over the phase factor $e^{i\theta(x)}$ and ensure all cancelations are correctly accounted for.

\begin{figure}[t]
\begin{center}
  \includegraphics[height=8cm]{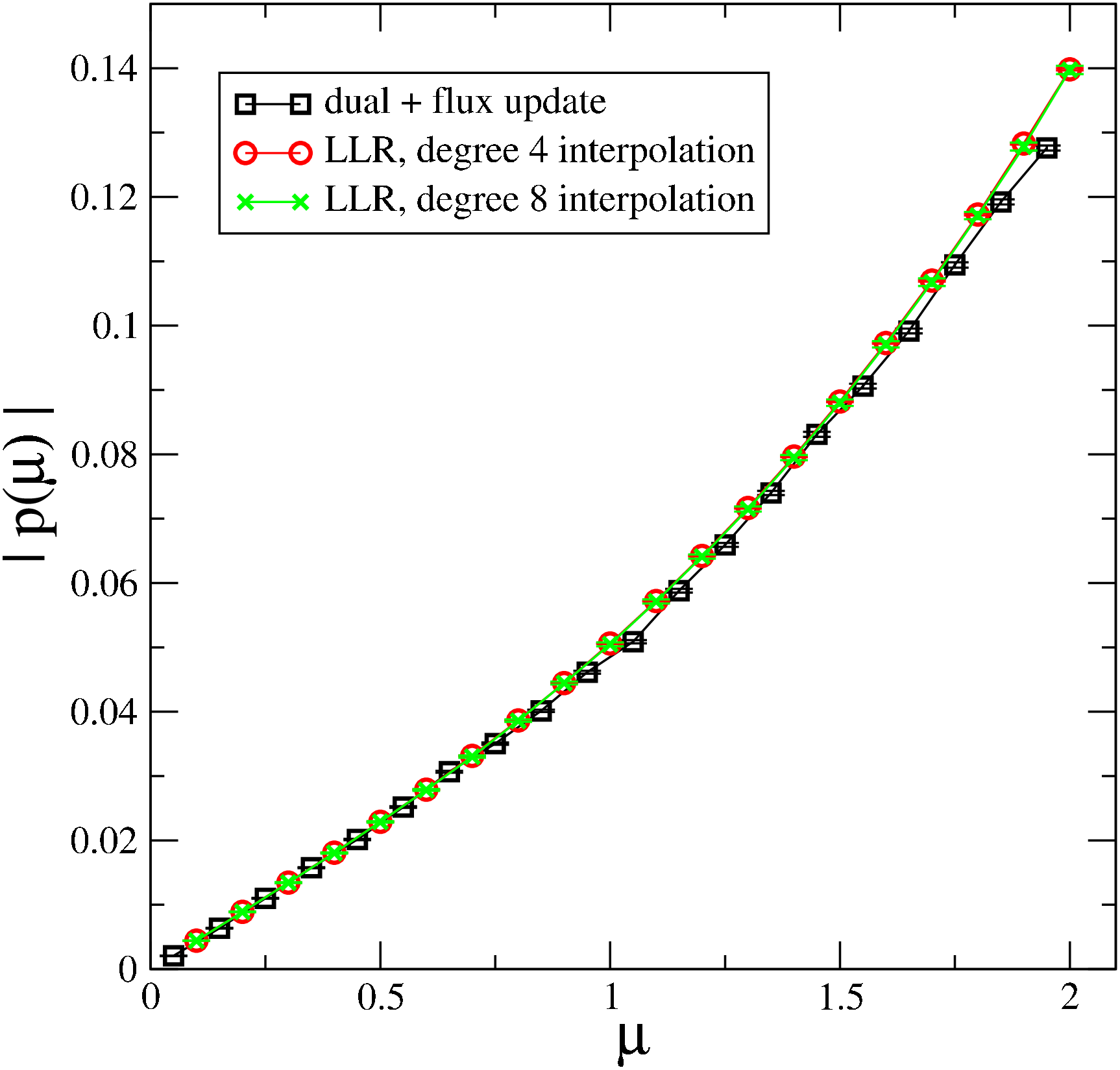}
\caption{Comparison between the results obtained with the density of states (LLR) and in the dual formulation in the Z(3) spin model at nonzero chemical potential.  From Ref.\ \cite{Lucini:2014wga}.}
\label{fig:dos2}
\end{center}
\end{figure}

In the Z(3) model it is possible to compare the results with an alternative dual formulation, which is manifestly free of the sign problem \cite{Mercado:2011ua}.  A comparison of the two approaches for the observable 
\be
|p(\mu)|\sim | \bra N_+-N_-\ket |
\ee
is shown in Fig.\ \ref{fig:dos2}, as a function of $\mu$. It is seen that the results from the dual formulation and the density of states agree, although there are still open questions about what happens at large $\mu$ and potentially at the transition \cite{Lucini:2014wga}.

In any case, it is clear that the LLR implementation of the density of states is a considerable improvement on previous histogram methods. Moreover, the extension to gauge theories is currently in progress.

\section{Complexification}
\label{sec:complex}

With a complex weight, it makes sense to look for dominant configurations in the path integral not on the original real manifold  but in the complexified configuration space. Indeed, the idea behind complex Langevin dynamics  \cite{parisi,klauder} is that there exists a real and positive distribution $P(x,y)$ (here for one real degree of freedom $x$), such that
\be
\int dx\, \rho(x) O(x)  = \int dxdy\, P(x,y) O(x+iy).
\ee
This distribution is effectively sampled by the stochastic process in $x$ and $y$. Equilibrium is reached as in Brownian motion.

For holomorphic actions, it has been established that the method is correct, provided that certain criteria for correctness are verified a posteriori \cite{Aarts:2009uq,Aarts:2011ax}. In nonabelian gauge theories, gauge cooling is essential to establish this \cite{Seiler:2012wz}. For meromorphic drifts, i.e.\ with poles, problems may arise, but not necessarily so \cite{Mollgaard:2013qra}.
The first results for full QCD were obtained by Sexty \cite{Sexty:2013ica}. Recent results for heavy dense QCD are summarised in Ref.\ \cite{Aarts:2014kja}. Some ideas based on experience with random matrix theory can be found in Refs.\ \cite{Mollgaard:2014mga,Splittorff:2014zca}.

\begin{figure}[t]
  \begin{center}
  \includegraphics[height=8cm]{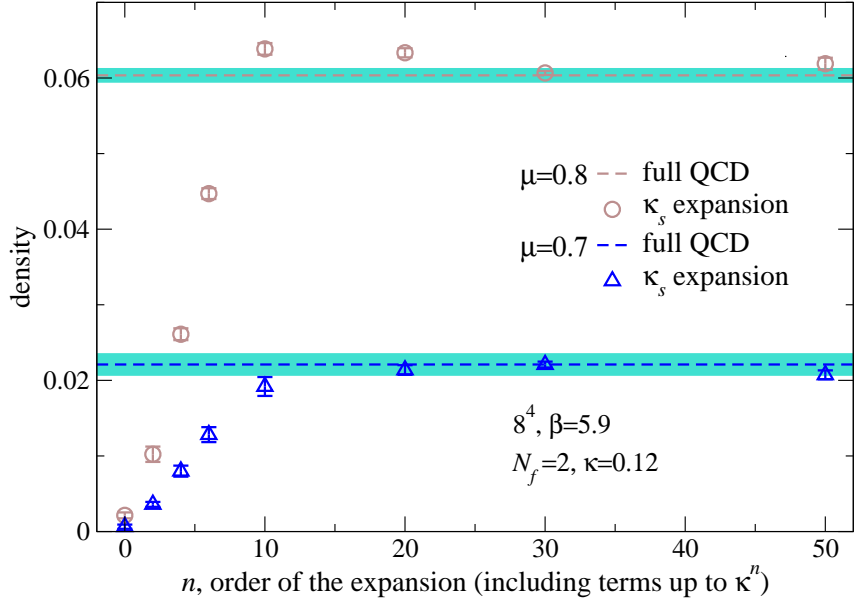}
  \caption{Dependence of the quark density (in lattice units) on the order of the truncation in the spatial hopping parameter ($\kappa_s$) expansion, for $\mu=0.7$ and 0.8, on a $8^4$ lattice. The lines show the result for full QCD. From Ref.\ \cite{Aarts:2014bwa}.}
  \label{fig:kappa}
  \end{center}
\end{figure}

A recent development in QCD is the implementation of the hopping parameter expansion to very high order, e.g.\ ${\cal O}(\kappa^{50})$, and a comparison with full QCD, i.e.\ without an expansion \cite{Aarts:2014bwa}. An illustrative result is shown in Fig.\ \ref{fig:kappa}, where the density is shown for two values of the chemical potential as a function of the order of the truncation in the hopping expansion. The hopping expansion used is an expansion in the spatial hopping parameter only ($\kappa_s$ expansion); the temporal hopping term is dealt with analytically. The truncated theory still suffers from a sign problem, which is handled with complex Langevin dynamics. 
Also shown are the results in full QCD, obtained with complex Langevin dynamics as well.
We observe therefore both a convergence of the hopping parameter expansion and an agreement with the full result, providing justification for both. More details were presented at this meeting by Sexty.

Another way to explore the complexified configuration space, possibly with more analytical control, is by integrating along so-called Lefschetz thimbles, i.e.\  using essentially a generalised saddle point expansion. In this setup one integrates numerically along lines of steepest descent, for which the imaginary part of the action in the weight is constant. Hence the sign problem is kept under control, at least to some extent (there is still a remaining residual sign problem, due to the curvature of the thimbles) \cite{Cristoforetti:2012su}. This method has now been implemented in various models and recent work focussed on an accurate computation of the residual phase \cite{Fujii:2013sra,Cristoforetti:2014gsa}. 
The most recent result is for chiral random matrix theory at nonzero chemical potential \cite{diR}.

\begin{figure}[t]
  \begin{center}
  \includegraphics[height=8cm]{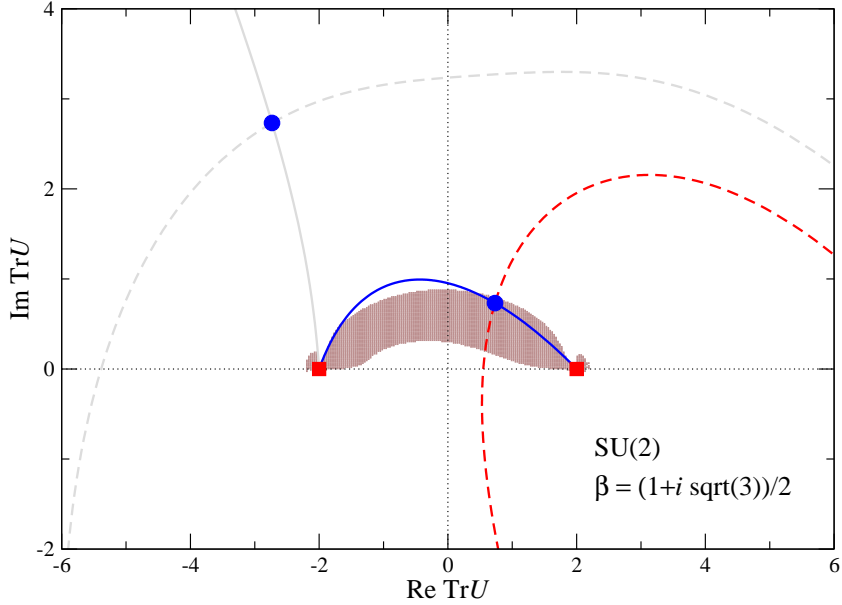}
 \caption{Comparison between the distribution sampled in complex Langevin dynamics (brown sausage shape)  and the stable/unstable thimbles (full/dashed) passing through the fixed points (blue circles) in the SU(2) one-link model with complex coupling $\beta=(1+i\sqrt{3})/2$. The red squares indicate where the reduced Haar measure has a zero and the thimbles end. From Ref.\ \cite{Aarts:2014nxa}.
   }
  \label{fig:su2}
  \end{center}
\end{figure}

Since both complex Langevin dynamics and the Lefschetz thimbles explore the complex field space, it makes sense to compare and contrast the two approaches. This has been done recently in Refs.\ \cite{Aarts:2013fpa,Aarts:2014nxa}, in the case of simple integrals of the form
\be
\label{eq:quartic}
Z = \int dx\, e^{-S(x)}, \quad\quad\quad S(x) = \frac{\sigma}{2}x^2 +\frac{\lambda}{4}x^4 + hx,
\ee
with $\sigma$ and/or $h$ complex, and for simple one-link U(1) and SU(2) models in the presence of a determinant and/or a complex coupling, for instance
\be
\label{eq:SU2}
Z = \int_{\rm SU(2)} dU \exp\left[\frac{\beta}{2}\Tr U\right],
\ee
with complex $\beta$ and a gauge symmetry, $U\to \Omega U\Omega^{-1}$, where $U, \Omega\in$ SU(2). 
 
For illustration we show in Fig.\ \ref{fig:su2} a result for the SU(2) model (\ref{eq:SU2}), with $\beta=(1+i\sqrt{3})/2$. The full (dashed) lines indicate the stable (unstable) thimbles, passing through the stationary points (blue dots). The red squares indicate a zero in the reduced Haar measure; here the thimbles end.
Integrating along the stable blue thimble yields the correct result. 
The distribution effectively sampled in the Langevin evolution is indicated with the shaded sausage-shaped region. Hence it can be observed that the thimble and the Langevin distribution cover a similar region in the complex configuration space around the attractive fixed point -- although the  thimble is one-dimensional and the latter two-dimensional. This observation appears to be generic: Langevin and thimble dynamics explore similar parts of the configuration space, although they differ in detail, e.g.\ due to the presence of repulsive fixed points in Langevin dynamics \cite{Aarts:2014nxa}.

\section{Outlook}

The sign problem is a fundamental obstacle in making progress towards the determination of the QCD phase diagram. Various ideas are currently being pursued and new algorithms are typically implemented in the simpler models first. For full QCD, complex Langevin dynamics appears to be the most promising, but caution and care have to be considered at all times.

\acknowledgments

It is a pleasure to thank Nucu Stamatescu, Erhard Seiler, D\'enes Sexty, Benjamin J\"ager, Lorenzo Bongiovanni and Felipe Attanasio for collaboration, Biagio Lucini, Francesco Di Renzo  and Kim Splittorff for discussion, and the organisers for an excellent meeting. This work is supported by STFC, the Royal Society, the Wolfson Foundation and the Leverhulme Trust. For computational resources, we thank the STFC funded DiRAC Facility and HPC Wales.

\end{document}